# The stiffness of elastomeric surfaces influences the mechanical properties of endothelial cells

Jagoba Iturri [1,*], Julia Miholich [1], Spela Zemljic [2], Amsatou Andorfer-Sarr [1], Rafael Benitez [3], and José L. Toca-Herrera [1,*]

[1] Institut für Biophysik, Universität für Bodenkultur Wien (BOKU), Muthgasse 11, 1190 Wien, Austria
[2] Department of Biophysics, Medicine Faculty, University of Ljubljana, Vrazov trg 2, 1000 Ljubljana, Slovenia
[3] Faculty of Economics, University of Valencia, Avda. Tarongers, s/n, 46022 Valencia, Spain

* Correspondence: jagoba.iturri@boku.ac.at; jose.toca-herrera@boku.ac.at

**Abstract:** Optimal characterization of the mechanical properties of both cells and their surrounding is an issue of major interest. Indeed, cell function and development are strongly influenced by external stimuli. Furthermore, a change in cell mechanics might, in some cases, associate with diseases or malfunctioning. In this work, atomic force microscopy (AFM) was applied to examine the mechanical properties of the silicone elastomer polydimethylsiloxane (PDMS) a common substrate in cell culture. Force spectroscopy analysis was done over different specimens of this elastomeric material containing varying ratios of resin to cross-linker in its structure (5:1, 10:1, 20:1, 30:1 and 50:1), which impacts the final material properties (e.g., stiffness, elasticity). To quantify the mechanical properties of the PDMS, factors as the Young's modulus, the maximum adhesive forces as well as both relaxation amplitudes and times upon constant height contact of the tip (in $t_{Dwell} \neq 0$) were calculated from the different segments forming the force curves. It is demonstrated that the material stiffness is increased by prior oxygen plasma treatment of the sample, required for hydrophilic switching, contrarily to what observed for its adhesiveness. Subsequent incubation of endothelial HUVEC cells on top of these plasma treated PDMS systems yields minor variation in cell mechanics in comparison to those obtained on a glass reference, on which cells show much higher spreading tendency and, by extension, a remarkable membrane hardening. Thus, surface wettability turns a factor of higher relevance than substrate stiffness inducing variations in the cell mechanics.

**Keywords:** Poly (dimethylsiloxane); Mechanical properties; Atomic Force Microscopy

## 1. Introduction

During their lifetime, cells experience a high number of external stimuli (physical forces, chemical signaling, etc.). Living cells are able to respond to these stimuli in different ways, and it is well known that their development and function are influenced by physical forces [1-3]. Moreover, it is known that some diseases, i.e. cancer, can be associated with changes in the mechanical properties of cells [4,5]. Carcinoma cells, for instance, show different elastic properties compared to healthy epithelial cells. The future use of cell biomechanics as a diagnostic tool might also be envisaged [6,7]. Then, it turns of great interest to characterize the mechanical properties of such biological samples. However, not only the mechanical properties of the cells themselves are of interest. It turns also important to characterize, from a mechanical point of view, the substrate used for cell/tissue culture known its influence onto cell mechanical properties and, consequently, cell growth and development [8-10].

Several experimental techniques, mainly based on the use of indenters, are usually employed to characterize mechanical properties of soft materials in both micro- and nanometric scale [11-14]. Among them, Atomic force microscopy (AFM) is a well-suited and popular technique to perform such indentation experiments at the nanoscale. In addition to its featuring performance as topography imaging tool it also allows measurements in Force Spectroscopy mode, by which the tip only moves along the vertical coordinate, perpendicularly to the sample of interest, under controlled motion speeds and maximum load upon contact with high accuracy in its location [15-17]. It offers several advantages compared to conventional nano-indenter instruments as, for instance, the on-demand choice of the indenter shape, size and stiffness, given through the use of different types of cantilevers as well as chemical modifications [18-21]. The data recorded, in the shape of Force vs distance curves, allows extracting various mechanical parameters attending to the segment of the curve chosen: Young's modulus and sample stiffness (approach-contact), maximum adhesion force (retract) and rheology-related stress relaxation and creep compliance –strain- (pause in contact) [15,16,22]. Combination of these complementary features ensures a rather complete characterization of the (bio)material of interest.

Here, Atomic Force Spectroscopy was used to firstly characterize mechanical properties of polydimethylsiloxane (PDMS), a popular silicon elastomer that, besides its application as e.g. implant or insulation material, it is widely used as a substrate for in vitro cell- and tissue culture [23-25]. Moreover, from a mechanical point of view, PDMS is more similar to tissue than other usual substrates (e.g. glass) and therefore a more natural scaffold for cell culture studies, especially for cell mechanics and cell spreading. Thus, the composition of PDMS (e.g., its mechanical properties) can be modified by controlling the elastomeric resin / crosslinker ratio during chemical synthesis [26,27].

Since substrate/cell interaction plays a crucial role in cell function and fate [28-31], proper characterization of these physico-chemical properties can become particularly useful for daily cell culture experiments. In this work, we have prepared PDMS films under different cross-linker compositions (5:1, 10:1, 20:1, 30:1, 50:1) as host substrate for HUVEC cells, as a model system to investigate substrate/cell interactions. We show how both the mechanical and wetting properties of the substrate influence cell spreading and its rheological properties.

**2. Materials and Methods**

    2.1.1    Preparation of PDMS films

In this work, Sylgard 184 Silicone elastomer kit (Dow Corning Corporation, U.S.) which consists of base (pre-polymer) and curing agent (cross-linker) was used for sample preparation. The crosslinking density and, therefore, the stiffness of the material can be influenced by varying the ratio of cross-linker to pre-polymer. To investigate PDMS of different composition, the following weight ratios (curing agent-to-base) were produced: 1:5, 1:10, 1:20, 1:30 and 1:50 whereby the 1:10 ratio is the composition proposed by the company.

First, the desired amount of curing agent was added to the pre-polymer with a Pasteur pipette and afterwards it was mixed properly with a spatula. Next, the PDMS was degassed for about 20–30 min in a vacuum desiccator to get rid of the air bubbles caused by mixing. After degassing, one droplet of PDMS mixture was spin coated onto a cover glass using a KLM Spin Coater (Schaefer Technologie GmbH). The spin coating procedure was performed at 2000 rpm for 60 s. Then, samples were cured in the oven (Memmert GmbH + Co.KG, Germany) for 3 h at 70 °C and kept in a petri dish at room temperature overnight.

    2.1.2    Plasma treatment

Samples were exposed to oxygen plasma for 20 s (60 W, flow rate of 90 $cm^3$ $min^{-1}$) using the PlasmaPrep2 (Gala Instrumente, Germany) instrument. This treatment, as mentioned above, affects the chemical structure of the top surface layer of the PDMS and causes a change in surface hydrophobicity (see section 2.1.1). To visualize the increase in hydrophilicity, the contact angle was measured with a FM40 EasyDrop instrument (KRÜSS GmbH, Germany). For that purpose, 10 µL of deionized water were put on the sample and the contact angle θ between both was determined

(data not shown here). However, because of the fact that this change is reversible [32], the AFM-measurements were carried out immediately after plasma treatment.

### 2.1.3 Cell culture

Human umbilical vein endothelial cells HUVEC (from ECACC). Cells were cultivated in DMEM GlutaMAX supplemented with 10% Fetal Bovine Serum (FBS) and 1 % penicillin/streptomycin at 37°C and 5% $CO_2$. For fluorescent staining, cells were cultivated until reaching 70-80% confluence. Then, HUVEC cells ($4\times10^4$) were seeded on functionalized glass slides and incubated for 24 hours at 37°C. After a washing step with 1x PBS, cells were incubated with Calcein-AM solution (1:100 dilution in PBS, stock: 1 mg/ml in DMSO) directly in the medium for 30 min at 37°C, followed by three washing steps with 1xPBS. The cells were incubated with 3 mL of the Hoechst 33342 staining solution (1:2000 dilution in PBS, stock: 10 mg/ml in water) for 5 min at 37°C. After washing three times with 1xPBS, Leibovitz L-15 medium was added, and cells were directly imaged by means of fluorescence microscopy

### 2.1.4 Force Spectroscopy

The force spectroscopy measurements on PDMS samples were performed on a JPK NanoWizard I (JPK TM Instruments AG, Germany). Uncoated silicon nitride cantilevers (NANOSENSORS Pointprobes, NanoWorld AG, Switzerland) with a nominal spring constant of 27 – 71 N m$^{-1}$ and a quadratic pyramid tip at the end were used. Cantilevers were cleaned prior to use with an UV Ozone ProCleaner (BioForce Nanosciences, U.S.) for approximately 20 min. The calibration of the spring constant was done with the thermal tune method. Individual force-time curves were recorded on at least two different samples on two positions each using an approach-speed of 0.4 μm s$^{-1}$. The maximum force applied was 1000 nN and the tip was kept in contact (in constant height mode) for 5 and 10 s. The measurements were conducted in air and in aqueous solution (ultrapure water).
Measurements on HUVEC cells, in turn, were performed on a NanoWizard III (JPK Instruments, Germany) with CellHesion, mounted on an inverted optical microscope (Axio Observer Z1, Zeiss). Experiments were done in Leibovitz L-15 medium at 37° C by using the JPK custom thermos-regulated flow-cell. The nominal spring constant of the employed Silicon-nitride probes (DNP-S, Bruker, USA) was of around 0.12 N m$^{-1}$. The use of a CellHesion system enables an extension of the Z height of the piezo up to 100 μm, required for cell experiments. Individual force-time curves were recorded on at least three different samples and 7 cells each using an approach-speed of 1 μm s$^{-1}$. The maximum force applied was set to 1.5 nN and the tip was kept in contact either in constant height or constant force modes for 5 and 10 s.

### 2.1.5 Data evaluation by R-software

Within this work, R-software [R Core Team, 2016] (version 3.2.3) was used for data evaluation of PDMS films. This is a software employed for statistical computing and batch processing. It also provides a broad range of graphical techniques so that several types of plots can be produced. Thus, it turns into a practical tool for calculation or manipulation of a large amount of data, as those generated from Force Spectroscopy studies. The functionality of R software can be extended/complemented by user-defined new functions or packages. For instance, in this work the data manipulation was done by means of afmToolkit package [33,34]. Such package provides various specific functions to analyze AFM force-distance curves.
Prior to their use into R environment, the raw data had to be converted into ASCII files by means of the JPK Data Processing Software, Version 4.2.62 (JPK Instruments AG, Germany). Once loaded, a function automatically detects the number of segments forming the force-distance curves (approach, contact, retract) and applies the corresponding algorithms for, among others, baseline correction and contact point determination. Indentation of the tip into the sample is calculated and after specifying parameters such as the Poisson-ratio, the elastic modulus is calculated according to the Hertz model (see section 2.1.6). This procedure can be done for various curves simultaneously and a table containing all values for E is finally obtained. Not only the evaluation of the Young's modulus but also the calculation of other properties like the data fitting for computing of the relaxation times and stress.

### 2.1.6 Mechanical parameters description

a) The Young's modulus (E), also called elastic modulus or modulus of elasticity, is the intrinsic parameter of the material related to its initial deformability. Thus, when a stress $\sigma$ is applied to an elastic material, the deformation induced (e.g., elongation) $\varepsilon$ is, according to Hooke's law, proportional to that stress.

Eq. 1 $$\sigma = E * \varepsilon$$

The proportionality constant *E* is defined as the elastic modulus or, in other words, the slope of the stress-strain curve in the elastic region. Hence, the higher the value for *E*, the stiffer the sample. In contrast, natural rubbers and silicones are common examples of elastomers showing viscoelastic properties. The response of such materials to stress is between that of a viscous material and an elastic material. Such materials can be then deformed elastically and characterized by an elastic modulus, but they also show viscous behavior, as described by some other parameters explained further below (see *Stress Relaxation* explanation). The Hertz model is one of the most commonly used models to calculate the Young's modulus *E* of a sample from an AFM force-distance curve [16] in which the contact between two linear, elastic spheres is described. However, this model also presents some limitations and does not consider adhesive forces, meaning that it is not applicable for sticky materials, and other method can be utilized. For example, the Johnson-Kendall-Roberts (JKR) theory [15] does consider the adhesive forces and can be combined with the Hertz model. Also, some adjustments considering other indenter geometries than spheres have been made in the past. Thus, for a tip shape of a four-sided pyramid, as used in this work, the following equation applies [35]:

Eq. 2 $$F = \frac{E * \delta^2 * \tan \alpha}{\sqrt{2}\,(1-\nu^2)}$$

Being $\alpha$ its face angle, the Poisson's ratio ($\nu$) a sample-dependent parameter usually set to 0.5 for incompressible materials like rubber and biological samples, and $\delta$ the indentation of the tip into the sample. Thus, for a pyramidal indenter the measured force is proportional to the square of the indentation.

To avoid substrate effects, the Hertz model is only applicable to the first 5 – 10% of the sample thickness. Since the spin-coated PDMS layer has a thickness of approximately 40 μm, consequently, only the first 2 -4 μm of indentation, from the approach segment, were considered for the fit (see Figure SI1, points b-c).

b) The maximum adhesive force ($F_{adh}$) parameter is, in turn, indicative of the stickiness of the film produced. It is brought by the minimum value of force recorded in the retraction segment of the force-distance plot. At such pulling position, the attractive forces between the tip employed and the sample under analysis reach their maximum. Afterwards, Force will then only decrease upon additional pulling, until tip-sample contact is totally lost (Figure SI1, point e and onwards).

c) Due to the viscoelastic nature of the sample, a phenomenon called *Stress relaxation* can in addition be observed, resulting from the "flowing" capability of the material [36] or, in other words, its rheological properties. Measuring of this factor is possible by keeping close contact between the tip and the material for a certain period of time, which is denoted as the Dwell time ($t_{Dwell}$). The contact is maintained right after the desired maximum setpoint value is achieved and takes place in constant height conditions of the cantilever. Hence, the Z position of the head is fixed, during the contact, to that reached at the setpoint. This allows the material to undergo structural rearrangement over the $t_{Dwell}$ of choice in response to the load-induced deformation (Figure SI2, points c-d).

## 3. Results and discussion

*3.1. Mechanical characterization of PDMS: Influence of cross-linker content*

3.1.1. Elastic modulus and adhesive force determination

An optimal characterization of elastomeric PDMS films by determination of their mechanical properties under different crosslinking degrees and surface wettability conditions, is of high utility in the context of biomedical applications, as aforementioned. Among these properties the Young's Modulus (E) and the maximum adhesion values can already describe, in a quite general manner, the response of such elastomer upon controlled loading and/or nano-indentation. In terms of Force Spectroscopy measurements, as defined for Atomic Force Microscopy technique, both E and maximum adhesion can be obtained from analysis of the force-distance plots recorded from either the approach or retraction segments, respectively (see Figure 1).

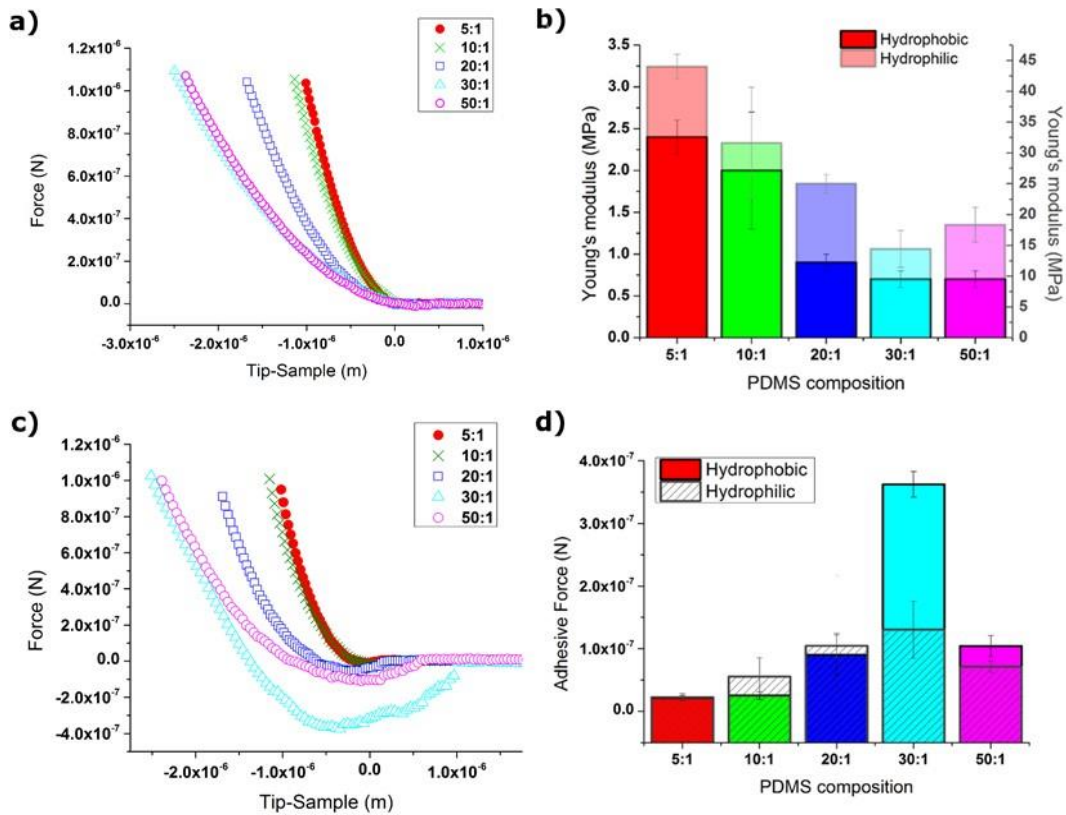

Figure 1. Force-distance curves comparing Approach vs Retract segment values. The stiffness (slope of the curve) of hydrophobic PDMS increases for large cross-liking values (Figure 1a). Note that for 1:30 and 1:50 ratios no significant variation could be measured. Figure 1b shows the values of the Young's modulus as a function of the PDMS composition (cross-linking). Note the linear dependence between E and the amount of curing agent (excluding the 1:50 ratio). Figure 1c relates to the retraction part of the force-distance curve. Both viscoelastic behavior and adhesive force (minimum of the plot), decrease for high cross-linking rations. The calculated adhesion forces are depicted in Figure 1d.

The amount of curing agent employed has a clear impact on the slope of the approaching force-distance curve (related to the stiffness of the sample) as shown in Figure 1a for fresh hydrophobic PDMS. Thus, the presence of lower amounts of crosslinker in the mixture (20:1, 30:1 and 50:1) led to a decrease in the stiffness of the film. For the 1:30 and 1:50 samples, the more viscous ones, no remarkable variation was observed. This might be due to the little amount of crosslinker present during the gel preparation (crosslinking procedure).

As explained in the Materials and Methods section, fitting of the initial 10% of the slope with a Hertz-derived model, adapted to the type of indenting tip employed, allows determining the Young's modulus of the sample. The values calculated are plotted in Figure 1b. A nearly linear dependence between the amount of curing agent and the E obtained can be observed for the untreated PDMS as well as for its plasma treated version, if the 1:50 ratio is excluded. The calculated values range between 0.7 and 2.4 MPa (wet conditions), for increasing crosslinker amount, which fit quite well with those given in the literature [25]. Again, the assumption that there is no real difference in deformability between the 1:30 and 1:50 PDMS is supported by fact that their respective approach curves show the same slope steepness. However, the general trend for the other systems shows how the more the crosslinking agent added to the sample, the higher the Young's modulus of the sample is. This is an expected observation and can be well explained by the fact that in the specimens containing less crosslinker, redundant polymer chains that were not crosslinked are present.

Table 1. Calculated Young's modulus values for PDMS specimens with different crosslinker ratios, and under different surface wettability and environmental conditions.

| | Young's Modulus (MPa) | | | | | | | |
|---|---|---|---|---|---|---|---|---|
| | Hydrophobic | | | | Hydrophilic | | | |
| Sample | Dry | SD | Wet | SD | Dry | SD | Wet | SD |
| 5:1 | 2 | 0.2 | 2.4 | 0.2 | 43.2 | 0.7 | 44 | 2 |
| 10:1 | 1.4 | 0.1 | 2 | 0.7 | 35.3 | 5.2 | 31.6 | 9.1 |
| 20:1 | 0.7 | 0.1 | 0.9 | 0.1 | 30.9 | 3 | 25 | 1.5 |
| 30:1 | 0.4 | 0.1 | 0.7 | 0.1 | 20.8 | 2.1 | 14.4 | 3 |
| 50:1 | 0.4 | 0.1 | 0.7 | 0.1 | 24.4 | 1.1 | 18.3 | 2.8 |

It is also important to note how a short plasma treatment (10 s) causes a 20 to 60-fold increase in the values of the elastic moduli, which shift to values between 14 and 44 MPa for minimum and maximum crosslinker contents, respectively. The finding of such an increased elastic modulus of PDMS after oxygen plasma treatment is consistent with the results presented in the work of Oláh et al [37]. The most plausible reason for that increase of stiffness is the formation of an oxidized surface layer induced by the oxygen plasma. An inorganic silica-like surface layer is formed that consists partly of silicon atoms bonded to 3 or 4 oxygen atoms ($SiO_x$, $1 \leq x \leq 2$). This silica-like structure is more rigid than the organic silicon from untreated elastomer films. Interestingly, hydrophilic specimens show a slightly smaller value for E when the measurement was performed in liquid environment compared to dry conditions (see Table 1 and Figure SI3). This observation may be rationalized as follows: The hydrophilic PDMS might be softened by water that is penetrating into the structure of PDMS. This could favor the migration of free siloxanes from the bulk to the surface through the porous silica-like layer what is believed to be the reason for the hydrophobic recovery of PDMS [37,38]. Contrarily no significant environment-derived changes could be observed on untreated hydrophobic PDMS during the first deformation steps. Due to the hydrophobic nature of the film, water is repelled from the sample and the contact remains unaltered compared to its dry version. However, since the seeding/culturing of cells must be performed on hydrophilic PDMS specimens, to ensure high cell-substrate affinity, only observations regarding this specific case turns of especial relevance for the second part of this work.

In turn, adhesion force values were determined from the f-d curves. As shown in Figure 1c, also the tip retraction segment of the corresponding force vs distance plots brings a variation like that of the approach motion, with crosslinker-dependent gradually varying slopes and zero-force recovery trends. Thus, for stiffer samples steeper decays and short recovery distances were measured, while the pulling range needed for full contact loss broadens on more viscous films. Such trend is observed again to occur for all the compositions measured except for the 50:1 ratio. Calculation of the maximum adhesion force (Figure 1d, Table 2) between our tip and the sample of interest, namely the minimum Y-axis position of the curve, corroborates the previous observations with values starting from around 8 nN for "stiff" 5:1 PDMS up to the 375 nN for the 30:1 composition, and then a drop to a fourth of it for high viscosity 50:1 films. Plasma-treated samples replicate such tendency in adhesion values, where the hardening of the outermost part of the sample seems to impact more severely PDMS films with low crosslinker content and causes a drop of around the 75% and 30% on 30:1 and 50:1 samples, respectively. However, on stiffer samples (10:1, 20:1) a contrary effect is observed, and adhesion values are even increased. This result is to some extent surprising by merely considering the glassy features originated on the interface through this treatment. An approximate explanation might derive from the appearance of small cracks on the surface of those structures, which provide a pathway for free PDMS migration, as already mentioned in literature [39]. The absence of this effect on the highest content of crosslinker, PDMS films in a 5:1 ratio, could then be due to the much lower number of free available chains within the structure compared to the 10:1 composition. Thus, for samples prepared in ratios 20:1 onwards the situation becomes paired because of their high fluidity and the adhesion measured stays in a similar level.

Table 2 Calculated adhesion force values for PDMS specimens with different crosslinker ratios, and under different surface wettabilities.

| Sample | Adhesive Force (N) | | | |
|---|---|---|---|---|
| | Hydrophobic | SD | Hydrophilic | SD |
| **5:1** | $2.16 \times 10^{-8}$ | $3.98 \times 10^{-9}$ | $2.27 \times 10^{-8}$ | $5.81 \times 10^{-9}$ |
| **10:1** | $2.54 \times 10^{-8}$ | $5.87 \times 10^{-9}$ | $5.55 \times 10^{-8}$ | $3.01 \times 10^{-8}$ |
| **20:1** | $9.03 \times 10^{-8}$ | $3.21 \times 10^{-8}$ | $1.04 \times 10^{-7}$ | $1.98 \times 10^{-8}$ |
| **30:1** | $3.62 \times 10^{-7}$ | $2.05 \times 10^{-8}$ | $1.30 \times 10^{-7}$ | $4.57 \times 10^{-8}$ |
| **50:1** | $1.04 \times 10^{-7}$ | $1.69 \times 10^{-8}$ | $7.13 \times 10^{-8}$ | $8.99 \times 10^{-9}$ |

3.1.2. Stress relaxation during contact ($t_{Dwell} \neq 0$)

Complementarily to those factors described above, the variation of crosslinking agent in the PDMS structure also influences the rheological properties of the sample. In this regard, a larger number of free polymer chains is clearly expected to confer a higher fluidity to the material. This, in extension, influences the way the polymeric material can react against application of distinct types of external forces. More specifically, in the case of a well-defined load on top of the film, as studied in this work, the capability of the underlying material to readapt structurally and to flow under such an interference can cause a damping of the stress applied. The time required by the different PDMS samples to relax an applied force of 1 µN (while keeping the height constant for $t_{Dwell}$ = 10 s) was determined by AFM. The results are shown in Figure 2a. Samples with the highest crosslinking degree (5:1 and 10:1) appear to behave almost identical, following comparable force decay paths, but as the resin/crosslinker ratio increases such relaxation process occurs in quite different ways. In turn, while 20:1 films render a large decay amplitude, the subsequent compositions (30:1 and 50:1) stay again out of the expected trend. The total force decay amplitudes measured at the end of the observation time are depicted in Figure 2b.

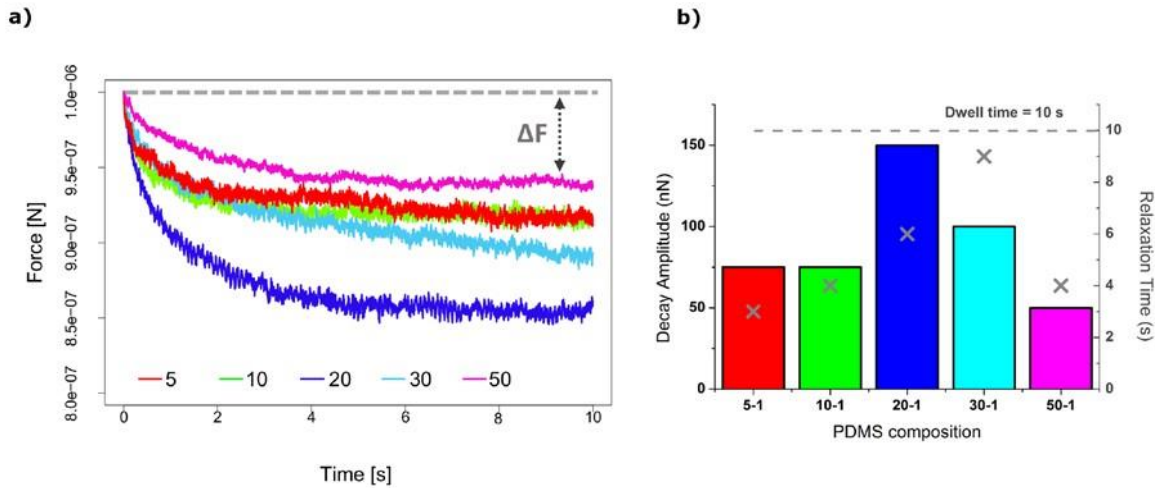

Figure 2. Stress relaxation experiment on PDMS films in different composition. Figure 2a shows how 310 PDMS samples relax after applying an external force of 1µN. The results indicate that the required 311-time scale is about 10 seconds for every sample. However, the decay amplitude depends on the cross312 linking. The relaxation times (X) and the decay amplitudes are depicted in Figure 2b.

Another interesting parameter to consider from these force vs distance plots would be the relaxation time ($t_{Relax}$) or, in other words, the time needed to achieve a plateau in the decay trend. This factor, divided by the observation time applied ($t_{Dwell}$), which has key influence on the achievement or not of the fully relaxed state, enables the calculation of dimensionless Deborah number [40], a value with mere rheological meaning and used to express the fluidity of the material.

An approximate value of $t_{Relax}$ can be easily extracted for each of the plots in Fig. 2a. For comparative purposes, as well as higher accuracy, this value could also be calculated by R software and then plotted in Fig. 2b. The relaxation time is a quick way to compare viscous vs elastic behavior (it is defined as the viscosity/Young´s modulus ratio). It can be seen how $t_{Relax}$ values increase, from around 3 seconds in a 5:1 system up to the 9 seconds required for 30:1 PDMS, by following an almost linear trend again interrupted by the drop shown for the 50:1 ratio.

In this case, it could also happen that in addition to the first plateau-like achieved, and due to the viscous nature of the film, the 50:1 PDMS could show a second stabilization point beyond the limits of the observation window. Such a comment would be motivated by the long relaxation measured for 30:1, already close the upper limits of the observation window. Interestingly, $t_{Relax}$ factor has also helped discriminating between 5:1 and 10:1 mixtures, which could not be done by measurement of their respective decay amplitudes.

*3.2. Mechanical characterization of HUVEC cells: glass vs PDMS*

Optimal characterization of the different substrates described above allowed addressing the study to a next level of complexity, in which HUVEC endothelial cells were seeded and cultured on top of those specimens, so the influence that the different compositions have in their mechanics could be analyzed. Plasma-treated glass substrates were employed as reference substrates, known their extended use and validity as cell culturing surfaces. In case of HUVEC cells, the study of their mechanics was limited to the determination of the corresponding Young's modulus value after 24 h of incubation. Force-distance curves were performed in an equivalent manner to that followed for PDMS substrates, choosing in this case the central position of the cell as indentation position.

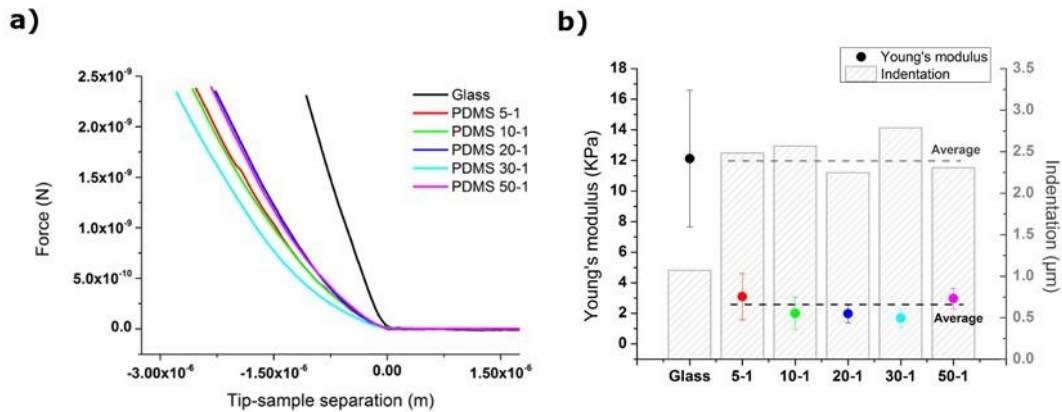

Figure 3. Mechanical properties of HUVEC cells on hydrophilic glass and different PDMS substrates. 348 The stiffness of HUVEC cells rises with increasing Young's modulus of the substrate (Figure 3a). The 349 values of the indentation (at constant applied load) and the Young´s moduli are shown in Figure 3b.

Figure 3a shows the average approach plots obtained for HUVEC on each of the underlying substrates attempted: hydrophilic borosilicate glass and PDMS in different crosslinking degrees. At a first glimpse, the force-distance plots from cells on glass show a much steeper evolution of forces upon indentation, indicating a much stiffer membrane than those of cells on PDMS, irrespectively to the crosslinking state. This result could be somehow predicted because of the higher affinity of cells for stiffer substrates (GPa vs MPa) under similar surface chemistry conditions of non-specificity. A more detailed analysis of PDMS systems reveals similar trends in the first 0.5-1 µm of indentation, with the exception of PDMS 30:1 substrates, whose force vs distance plot appears steeper than it theoretically should. The Hertz model was used to fit the experimental curves to obtain the Young´s modulus (E). Figure 3b shows the values of E as well as the indentation required for maximum setpoint/load achievement. It can be clearly observed how Young's modulus of HUVEC on glass (12 MPa) results about 4-5 times higher than on PDMS samples but, in turn, also the standard deviation is larger. Elastic moduli values of cells seeded for 24 hours on plasma-treated PDMS are 3.20, 2.0, 1.98 1.69 and 2.97 MPa, if ordered in decreasing amount of crosslinker in the substrate underneath. Thus, the cell stiffness variation is not as evident as expected, despite the gradual (and logical) decrease observed between 5:1 and 30:1 systems. Similar conclusions derive from the analysis of the indentation required for achieving the maximum load: for HUVEC cells on glass the setpoint is reached in around 1 µm, less than a half of the average indentation needed on PDMS-supported cells, waving between 2.2 and 2.7 µm. Hence, the assumption that the softer the material, the lower the cell-substrate affinity and, by extension, the rounder and softer the cell will appear is here only partially covered.

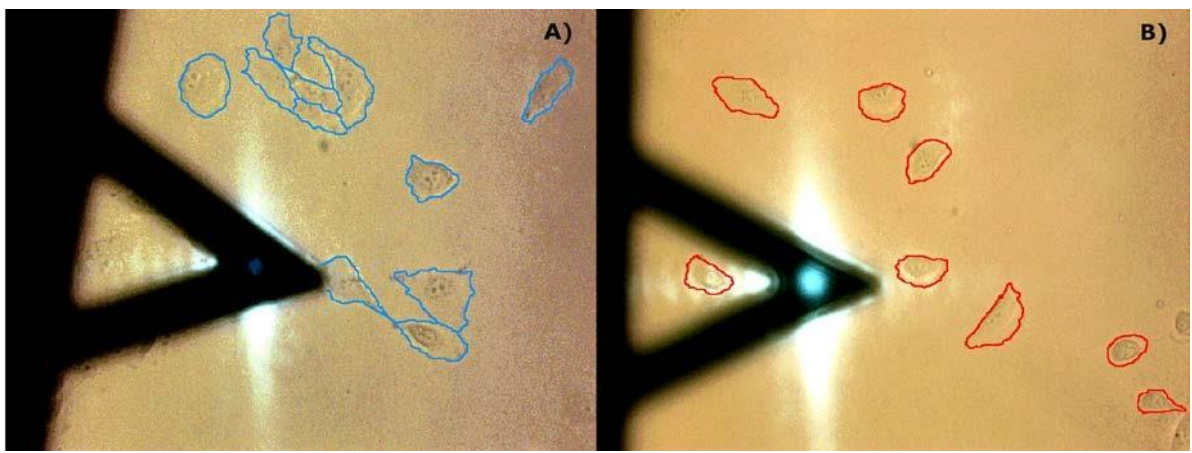

**F**igure 4. Optical Micrographs (x20) of HUVEC cells on glass (a) and PDMS 5:1 (b) after 24 hours of incubation. Cell profiles are highlighted in blue and red, respectively, for a better visualization of the area occupied by cells in their spreading.

Such statement could only be fully corroborated, for instance, when comparing the shape of HUVEC cells cultured either on glass or 5:1 PDMS, where the stiffness featuring each of the materials differs in one order of magnitude, instead of only a few MPa, as the optical microscopy images reveal in Figure 4. The profiles of the cells are highlighted in blue and red for glass and PDMS substrates, respectively, for the sake of a better visualization. Here, two main observations arise from a quick comparison between both images: (a) HUVEC cells on glass tend to cover larger areas due to their better spreading, which induces membrane hardening and originates their higher E values, and (b) cells on glass are more prone to forming tissue-like assemblies in comparison to isolated cell appearance on PDMS, independently of their composition. This reinforces the concept of needing a large variation (here of an order of magnitude) between the mechanical parameters of the substrates employed for HUVEC seeding, to impact their mechanical behavior. The fine tuning of the crosslinking degree seems not to be sufficient to induce large structural arrangements in cells despite the plasma treatment applied prior to cell culturing.

**Acknowledgements:** Alberto Moreno-Cencerrado is acknowledged for technical support. This work was partially supported by the International Graduate School BioNanoTech (IGS) Program of the Federal Ministry for Science and Research (Austria), and the U.S. Air Force Office of Scientific Research (AFOSR) Projects FA9550-07-0313 and FA9550-09-1-0342.

**Author Contributions:** J.I., S. Z. and J.-L. T.-H. conceived and designed the experiments; J. I. and J. M. performed the experiments; J.I. and J.M. analyzed the data; S. Z. and A. A.-S. provided biological materials and took care of cell culture; R. B. wrote the R package for data analysis, J.I. and J.-L. T.-H. wrote the paper.

**Conflicts of Interest:** The authors declare no conflict of interest. The founding sponsors had no role in the design of the study; in the collection, analyses, or interpretation of data; in the writing of the manuscript, and in the decision to publish the results.

## Supplementary information

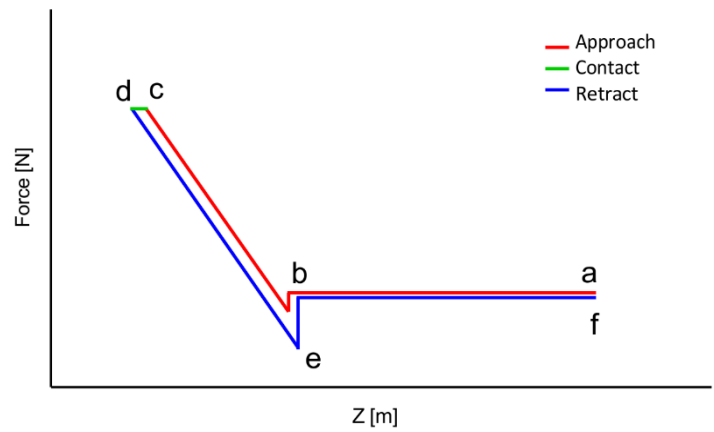

**Figure SI1.** Schematic drawing of a force-distance curve. The cantilever is approached from the starting point (a) towards the sample and jumps to contact at point (b) when it is close enough to the surface. It is then deflected because it is pressed into the sample until a predefined setpoint is reached (c). After the contact segment going from (c) to (d), dependent on the Dwell time chosen, the cantilever is retracted again from the sample and it leaves the surface of the sample, measuring adhesion events (e). Finally, it moves to the initial position (f) again in the absence of any interaction (coming back to it rest position, zero force).

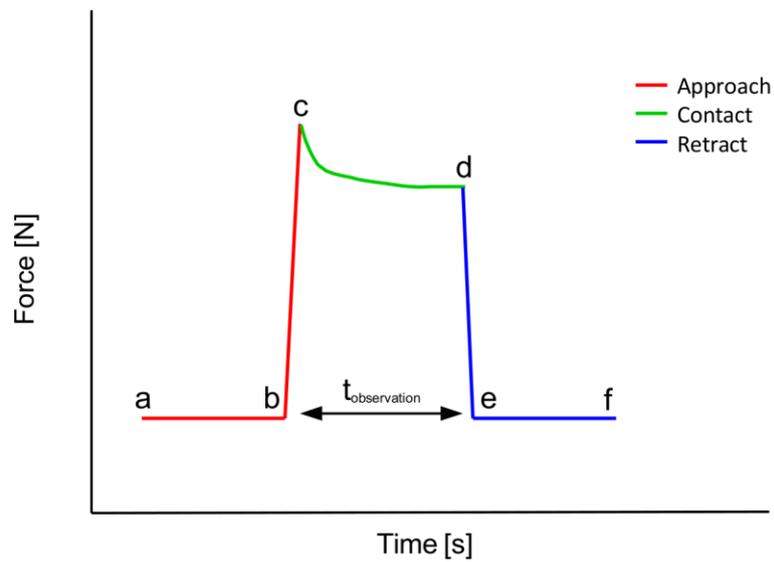

**Figure SI2**. Constant height experiment. The typical force vs. time plot is depicted for the three different segments, as shown by the colored lines. Definition of each point is the same as in Figure SI1.

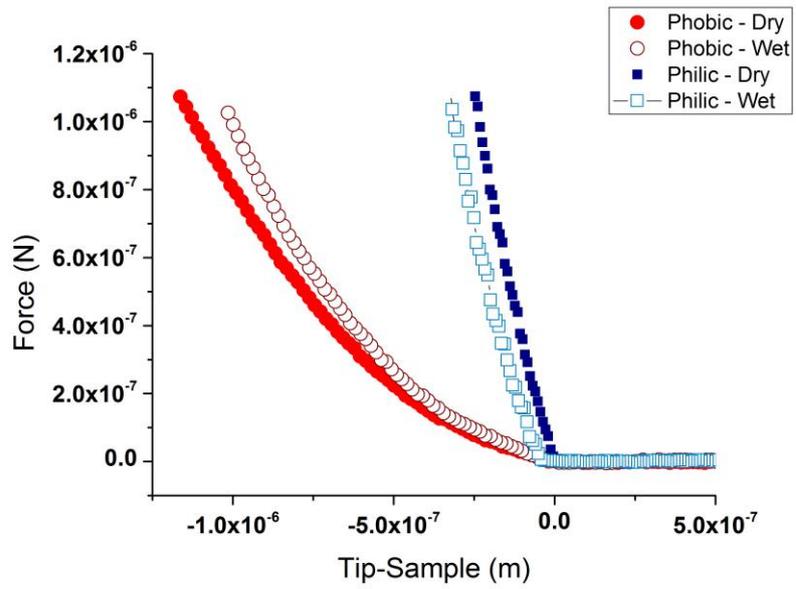

**Figure SI3.** PDMS 10:1 Dry vs Wet approach segment comparative plots. Note that water does not influence the tendency of the stiffness of the sample (slope of the force-distance curves). Hydrophilic samples are stiffer than hydrophobic ones.